\documentclass[final]{aipproc}

\layoutstyle{6x9}

\begin{document}

\title{Metastability and Avalanches in a Nonequilibrium Ferromagnetic System}

\author{Pablo I. Hurtado}
{
  address={Institute 'Carlos I' for Theoretical and Computational Physics, 
and Departamento de Electromagnetismo y F{\'\i}sica de la Materia,
University of Granada, 18071, Granada, Spain.}
}

\author{J. Marro}
{
  address={Institute 'Carlos I' for Theoretical and Computational Physics, 
and Departamento de Electromagnetismo y F{\'\i}sica de la Materia,
University of Granada, 18071, Granada, Spain.}
}

\author{Pedro L. Garrido}
{address={Institute 'Carlos I' for Theoretical and Computational Physics, 
and Departamento de Electromagnetismo y F{\'\i}sica de la Materia,
University of Granada, 18071, Granada, Spain.}}

\begin{abstract}
We present preliminary results on the metastable behavior of a nonequilibrium ferromagnetic system.
The metastable state mean lifetime is a non-monotonous function
of temperature; it shows a maximum at certain non-zero temperature which depends on the strengh
of the nonequilibrium perturbation. This is in contrast with the equilibrium case in which lifetime  
increases monotonously as the temperature is decreasesed. We also report on avalanches during the
decay from the metastable state. Assuming both free boundaries and nonequilibrium
impurities, the avalanches exhibit power-law size and lifetime distributions. Such scale free behavior is 
very sensible. The chances are that our observations may be observable in real (i.e. impure) ferromagnetic 
nanoparticles.
\end{abstract}

\maketitle


Metastability is ubiquitous in Nature, and it often determines the system behavior. 
The microscopic understanding of this phenomenon, which is mathematically
challenging,\cite{Pnros} is therefore of great practical and theoretical interest. 
In particular, metastability is relevant to the behavior of magnetic storage devices.
A magnetic material may consist of 
magnetic monodomains. In order to store information on such material, one magnetizes
each individual domain using a strong magnetic field, defining in this way a bit of information. 
A main concern is to retain the individual orientations
of the domains for as long as possible in the presence of weak arbitrarily-oriented 
external magnetic fields. The interaction 
with these external fields often produces metastable states in the domains, and 
the resistance of stored information thus depends on the properties of these metastable 
states, including the details of their decay.
On the other hand, real magnetic domains are usually {\it impure}.
The microscopic nature of impurities,  which shows up in actual specimens
as spin, bond and/or lattice disorder and other inhomogeneities, quantum
tunneling,\cite{Vacas} etc., suggests they might dominate the behavior of
near$-$microscopic particles; in fact, they are known to influence even
macroscopic systems. An interesting issue is therefore understanding the formation 
of a new phase inside a metastable domain which contains impurities.

Following recent efforts,\cite{efforts} we study in this paper the  simplest possible model of this 
situation, namely a $2d$ Ising ferromagnet with periodic boundary conditions that we 
endow with a weak dynamic perturbation competing with the thermal spin-flip
process, which impedes equilibrium. 
Consider the Hamiltonian function
$\mathcal{H}(\vec{s})  =-J\sum_{\langle i,j \rangle}s_{i}s_{j}-h\sum_{i=1}^{N}s_{i}$,
where $J>0$ (ferromagnetic interactions), $s_{i}=\pm 1$ stands for the two
possible states of the spin at site $i$ of the square lattice, $i=1,...,N$,
and the first sum is over any pair $\langle i,j \rangle$ of
nearest$-$neighbor sites. The lattice side is $L$; $N=L^2$.
The system configuration, $\vec{s} \equiv \{s_{i}\}$, 
is let to evolve in time due to superposition of two
canonical drives. That is, we chose the transition probability per unit time
for a change of $\vec{s}$ to be
\begin{equation}
\omega(\vec{s} \rightarrow \vec{s}^{\ i})  =p+(1-p)
\frac{\textrm{e}^{-\frac{1}{T}\Delta\mathcal{H}_{i}}}{1+\textrm{e}
^{-\frac{1}{T}\Delta\mathcal{H}_{i}}} 
\label{eq102}
\end{equation}
Here $\vec{s}^{\ i}$ stands for $\vec{s}$ after flipping the spin at $i$, 
and $\Delta\mathcal{H}_{i}\equiv\mathcal{H}(\vec{s}^{\ i})  -\mathcal{H}(  \vec{s})$. The 
Boltzmann constant is set $k_{B}\equiv 1$ in this paper.
One may interpret that rule (\ref{eq102}) describes a spin$-$flip mechanism
under the action of two competing heat baths: with probability $2p$, the flip
is performed completely at random (as if $\vec{s}$ were in contact with a heat
bath at `infinite' temperature), while the change is performed at temperature
$T$ with probability $1-p$. For $p=0$, eq. (\ref{eq102}) 
corresponds to the equilibrium Ising case, which exhibits for $h=0$ a critical point 
at $T=T_{C}\approx 2.2691\,J$. Otherwise, the conflict in (\ref{eq102}) 
impedes canonical equilibrium, and the system evolves with time 
towards a non$-$equilibrium steady state that essentially differs
from the Gibbs state for $T$. It is assumed that this kind of stochastic,
non$-$canonical perturbation for $p>0$ may actually occur in Nature due to
microscopic disorder or impurities, etc.\cite{Vacas} 
The system shows for $p\neq 0$ and $h=0$
a continuous order-disorder phase transition at a critical temperature $T_c(p)$. This critical
temperature can be calculated in first order mean field approximation as\cite{Pablotesis}
\begin{figure}
\includegraphics[height=.3\textheight]{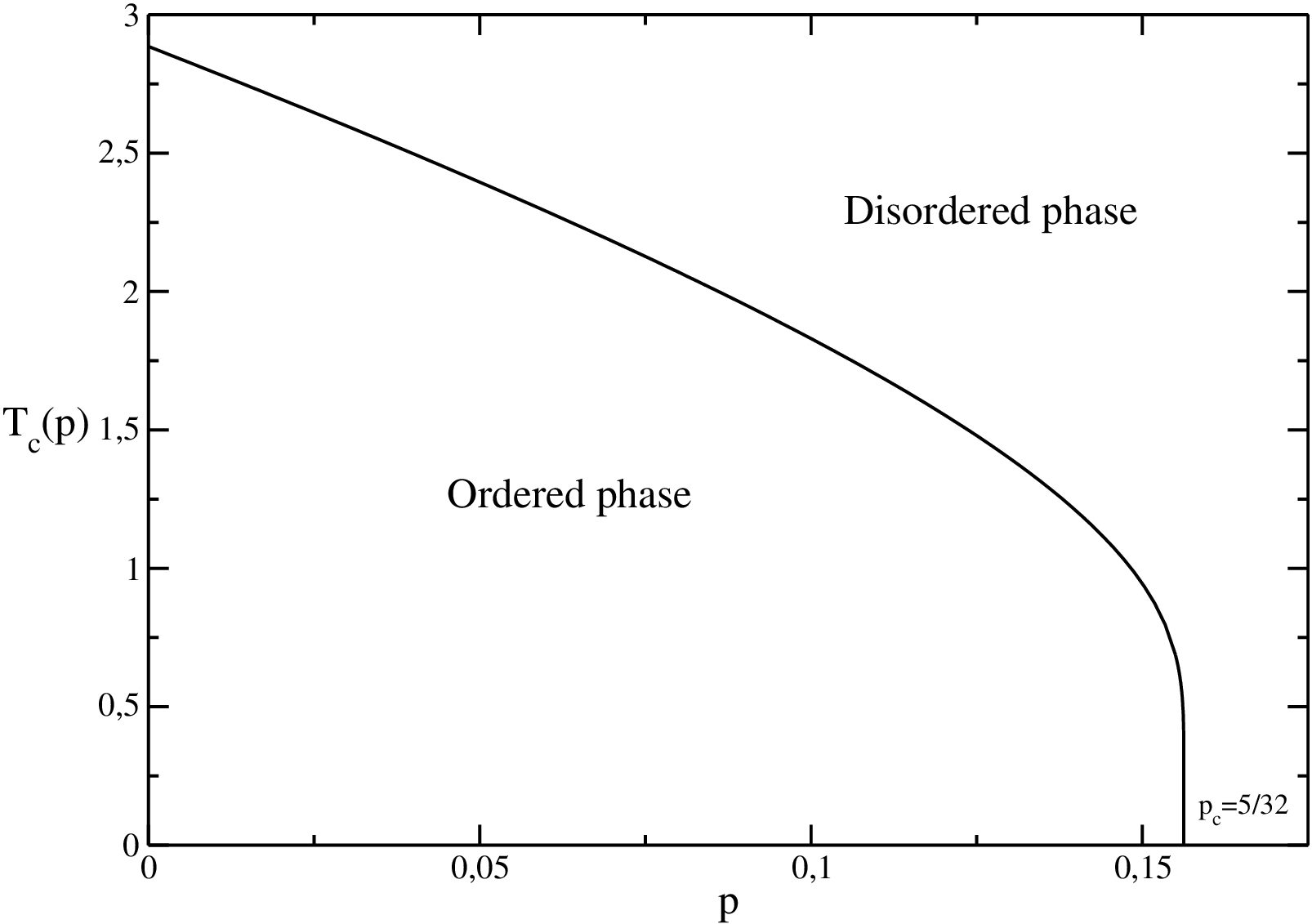}
\caption{\small Critical temperature for our nonequilibrium ferromagnetic system as a function of $p$ in 
a first-order mean-field approximation.}
\label{diagfases}
\end{figure}
\begin{equation}
\frac{T_c(p)}{J}=\frac{-4}{\ln [-\frac{1}{2} + \frac{3}{4} \sqrt{\frac{1-4p}{1-p}}]}
\label{tempBethe}
\end{equation}
Fig. \ref{diagfases} shows $T_c(p)$ as a function of $p$. 
$T_c(p=0)$ is the Bethe temperature, $T_{Bethe} \approx 2.8854\,J$, to be compared with the 
exact critical value for $p=0$, $T_C$. There is a critical value of $p$, $p_c$, such that for larger values 
of $p$ there is no ordered phase for any temperature.  This can be obtained from the 
condition $T_c(p_c)=0$, yielding $p_c=\frac{5}{32}=0.15625$. In this paper we pay attention to the ordered phase.

\begin{figure}
\includegraphics[height=.3\textheight]{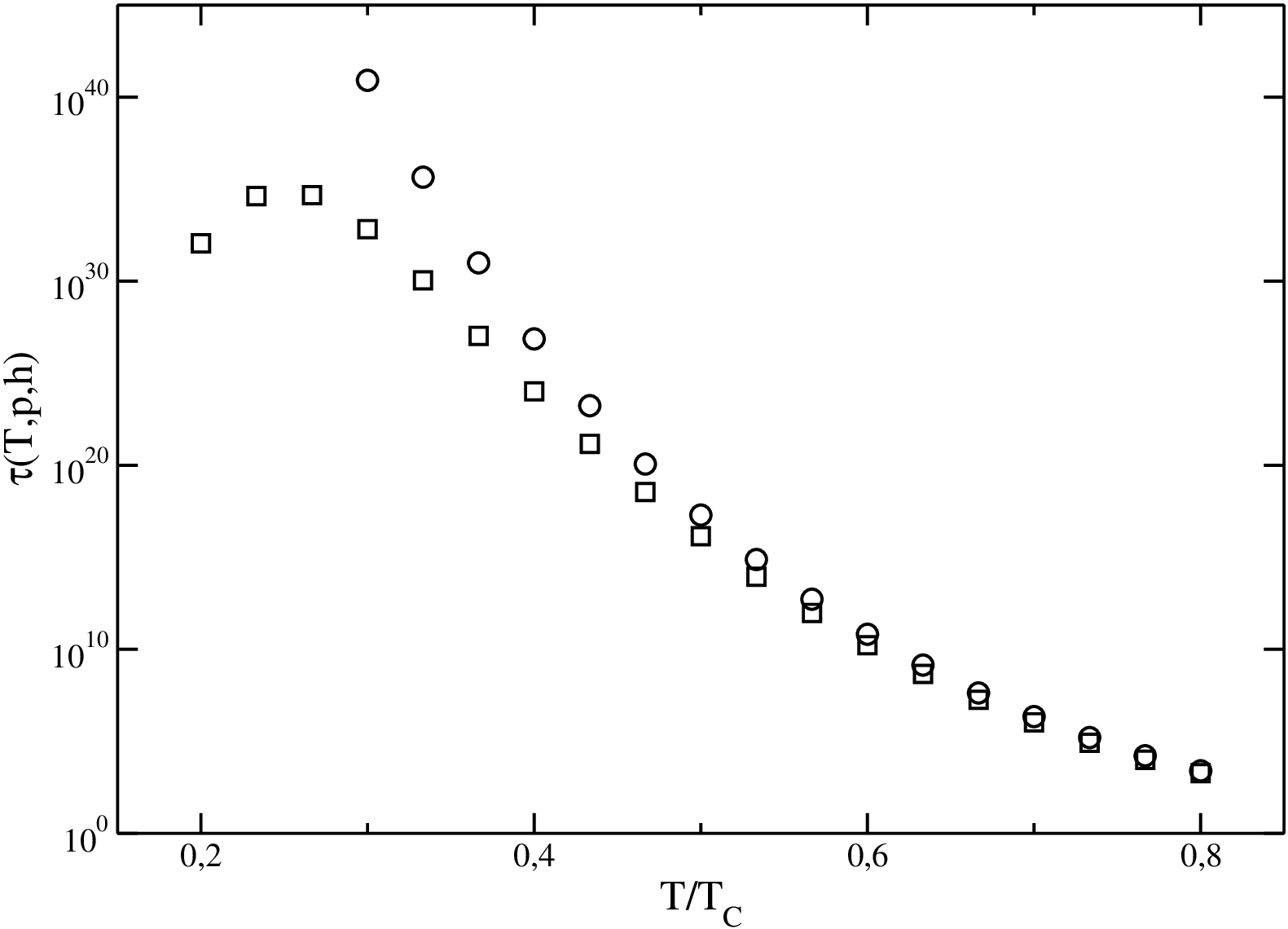}
\caption{\small Semilog plot of the metastable-state mean lifetime as a function of temperature for a 
system with side $L=53$ and, from top to bottom, $p=0$ (equilibrium) and $p=0.001$ (nonequilibrium). The data come from 
an average over $1000$ independent runs. Notice the non-monotonous behavior of $\tau$ for the nonequilibrium case.}
\label{vida}
\end{figure}

In order to characterize the metastable behavior of the model,
we measured the mean lifetime for values of $T$ and $p$ such that $T<T_c(p)$ and a magnetic field $h=-0.1$. 
We take the initial state to be $s_i=+1$ for $i=1,\ldots,N$. Under
the negative field, this ordered state is metastable, and it eventually decays to the stable state. 
In fact, the system rapidly evolves from the initial state to a state in the metastable region, with magnetization 
close to $+1$. After this fast relaxation, the system spends a long time wandering around the metastable state,
eventually nucleating one or several critical droplets of the stable phase, which rapidly grow thus making 
the system to evolve from the metastable state to the stable one. We define the mean lifetime of 
the metastable state, $\tau(T,p,h)$, as the mean first-passage time (in Monte Carlo steps per spin, MCSS) to $m=0$.
The simulations reported here required in practice using the {\it s-1 Monte Carlo with absorbing 
Markov chains} (MCAMC) algorithm, and the {\it slow forcing approximation}.\cite{MCAMC}
Fig. \ref{vida} shows $\tau(T,p,h)$ as a function of T for a $L=53$ and $p=0$ (the equilibrium case) and 
$p=0.001$. Rather amazing, we observe in the figure how the nonequilibrium lifetime exhibits a maximum and then decreases
as the temperature drops. The maximum occurs at $T_{max}(p)$ which depends on the nonequilibrium parameter $p$. 
On the contrary, the lifetime in the equilibrium case grows exponentially fast as a function of $1/T$, as predicted by 
nucleation theory\cite{efforts}. We do not have a simple explanation of this nonequilibrium effect.
However, it may be emphasized that it has some practical implications. That is, one should not blindly decrease temperature
when trying to maximize the lifetime of real (i.e., impure) metastable magnetic particles, but look for the 
temperature $T_{max}(p)$ which maximizes the lifetime for the typical impurity concentration of the particle.

The previous observations concern the {\it bulk} metastability. However, in real magnets, one needs in practice to 
create and to control fine grains, i.e., magnetic particles with {\it borders} whose size ranges from mesoscopic 
to atomic levels, namely, clusters of $10^{4}$ to $10^{2}$ spins, and even smaller ones. Though 
experimental techniques are already accurate for the purpose,\cite{Shi} the underlying
physics is much less understood than for bulk properties. In particular, 
one cannot assume that such particles are neither \textit{infinite} nor
\textit{pure.} That is, they have free borders, which results in a large
surface/volume ratio inducing strong border effects, and impurities.
Motivated by the experimental situation, we studied a finite,
relatively small two$-$dimensional system subject to open circular boundary conditions. 
The system is defined on a square lattice, where we inscribe a circle of radius $R$; 
sites outside this circle do not belong to the system and are set $s_i=0$. 
We mainly report in the following on a set of fixed values for the model parameters,
namely, $h=-0.1$, $T=0.11T_{C}$ and $p=10^{-6}$. 
This choice is dictated by simplicity and also because (after
exploring the behavior for other cases) we came to the conclusion that this
corresponds to an interesting region of the system parameter space, where the
effects of $p$ and $T$ are comparable and clusters are compact and hence easy to analize.
We believe that we are describing here typical behavior of our model, and the chances 
are that it can be observed in actual materials.

The effects of free borders  on the metastable-stable transition have already been
studied for equilibrium systems.\cite{contorno} In this case, the system evolves 
to the stable state through the {\it heterogeneous} nucleation of one or several critical
droplets which always appear at the system's border. That is, the free border acts as a droplet condenser. 
This is so because it is energetically favorable for the droplet to nucleate at the border. Apart from the observed 
heterogeneous nucleation, the properties of the metastable-stable transition in equilibrium ferromagnetic nanoparticles 
do not change qualitatively as compared to the periodic boundary conditions case.\cite{contorno} In our nonequilibrium
system we observe a similar behavior, namely heterogeneous nucleation and the same qualitative nucleation properties.
However, the fluctuations or noise that the nonequilibrium metastable system shows as 
it evolves towards the stable state subject to the combined action of free borders and the nonequilibrium perturbation
are quite unexpected.
\begin{figure}[t]
\includegraphics[height=.24\textheight]{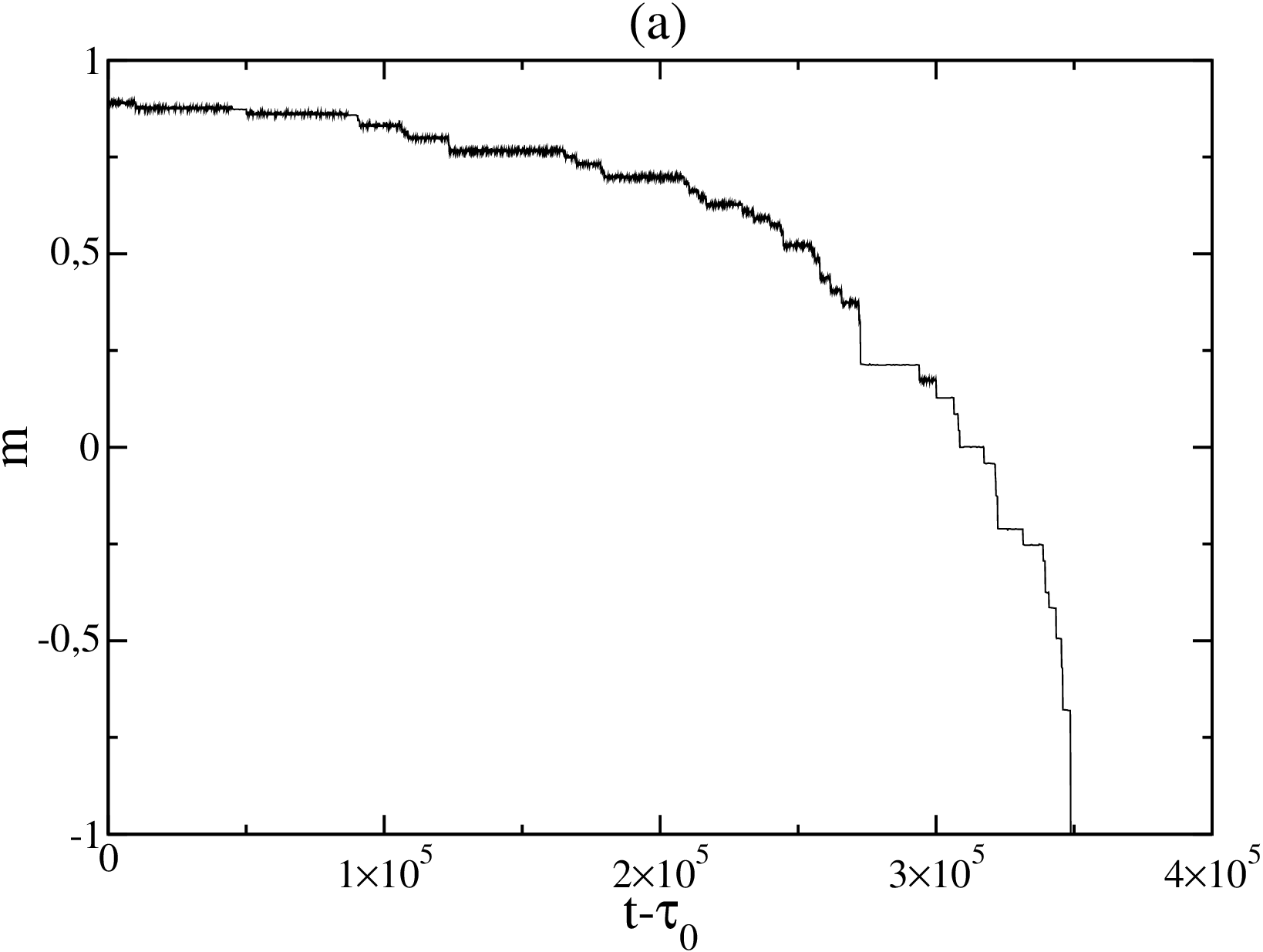}
\includegraphics[height=.24\textheight]{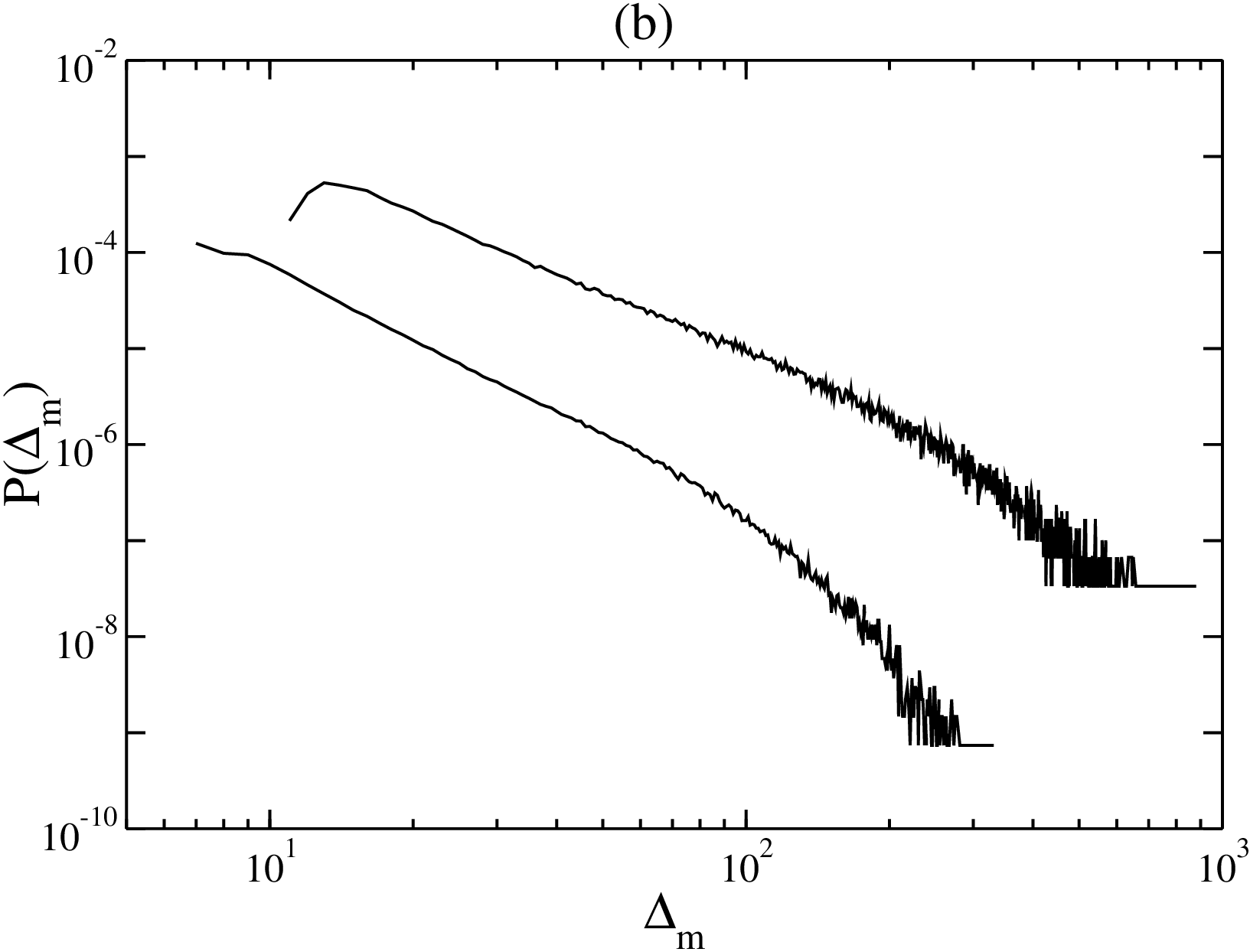}
\caption{\small (a) Time variation of the magnetization showing the decay from a metastable state for a $R=30$ particle; 
avalanches are seen here by direct inspection. Time is in Monte Carlo Steps per Spin (MCSS), and $\tau_0\sim 10^{30}$ MCSS.
(b) Large avalanche size distribution $P(\Delta_m)$ for the circular magnetic nanoparticle and $R=30$ 
(bottom) and $R=60$ (top). The second curve has been shifted in the 
vertical direction for visual convenience.}
\label{evolaval}
\end{figure}

As illustrated in Fig. \ref{evolaval}.a, the relaxation of magnetization occurs via
a sequence of well$-$defined abrupt jumps. That is, when the system relaxation
is observed after each MCSS, which corresponds to a 'macroscopic' time scale,
{\it stricktly monotonic changes of $m(t)$} can be identified that we shall
call \textit{avalanches} in the following. 
To be precise, consider the avalanche beginning at time $t_{a}$, when the system
magnetization is $m(t_{a})$, and finishing at $t_{b}$. We
define its \textit{size} and \textit{lifetime} or \textit{duration},
respectively, as $\Delta_{m}= | m(t_{b})-m(t_{a}) |$ and
$\Delta_{t}= | t_{b}-t_{a} |$. Our interest is on the histograms 
$P(\Delta_m)$ and $P(\Delta_t)$. Fig. \ref{evolaval}.b shows the large avalanche size distribution
$P(\Delta_m)$ for particle sizes $R=30$ and $R=60$, after the extrinsic noise\cite{Spaso} (i.e. the trivial, 
exponentially distributed small avalanches) has been substracted. A power law behavior, followed by a
cutoff is clearly observed. The measured power law exponents, $P(\Delta_m) \sim \Delta_m^{-\tau (R)}$,  
show size-dependent corrections to scaling. Similar corrections have been also found in real experimental 
systems.\cite{Frette} In particular we find $\tau (R=30)=2.76(2)$ and $\tau (R=60)=2.06(2)$. The 
lifetime distribution also shows power law behavior, $P(\Delta_t) \sim \Delta_t^{-\alpha (R)}$ with a
cutoff. Here we measure $\alpha(R=30)=3.70(2)$ and $\alpha(R=60)=2.85(2)$. This power-law behavior implies that
avalanches are scale-free (up to certain maximum size and lifetime) in our nonequilibrium ferromagnet subjected to
open boundary conditions. We also measured avalanches for $p=0$ in the circular magnetic particle case,
and for $p\neq 0$ in the periodic boundary conditions system. In both cases only small avalanches occur and the distributions 
are exponential-like, thus indicating the absence of scale invariance. That is, the combined action of free boundaries and 
impurities is behind the large, scale-free avalanches and essentially differs from the
standard bulk noise driving the system and causing small, exponentially distributed avalanches only. The physical origin
of this scale invariant behavior will be studied in a forthcoming paper.

Summing up, in this paper we present preliminar results on the metastable behavior of a nonequilibrium ferromagnetic
system. The presence of nonequilibrium conditions considerably enriches the observed phenomenology. In particular,
we study the metastable-state mean lifetime for a lattice with periodic boundary conditions.
Under the action of the nonequilibrium perturbation parametrized by $p$, the lifetime $\tau(T,p,h)$ shows a maximum
as a function of $T$ for certain nonzero temperature $T_{max}(p)$, then decreasing for lower temperatures. 
This counter-intuitive behavior, not observed in equilibrium, has some practical implications for real
magnetic systems with {\it impure} ferromagnetic domains. We also observe that, under the action
of both the nonequilibrium impurity and free borders, the metastable-stable transition proceeds by avalanches. These
are power-law distributed, thus showing scale invariance (up to certain cutoffs). 
The chances are that our observations about the effect of the nonequilibrium conditions on the properties of metastable states
and their decay, which we can prove in our model cases, are a general feature of similar phenomena in real
magnetic domains.

\begin{theacknowledgments}
We acknowledge M.A. Mu\~noz and S. Zapperi for very useful comments, and M.A. Novotny for sharing unpublished 
information about MCAMC algorithms. We also aknowledge financial support from Ministerio de Ciencia y 
Tecnolog\'{\i}a$-$FEDER, project BFM2001-2841.
\end{theacknowledgments}

\end{document}